\documentclass[onecolumn,floatfix,superscriptaddress,showpacs,showkeys,nofootinbib,preprint]{revtex4} 

\usepackage{epsfig}
\usepackage{amssymb,latexsym,amsmath}
\newcommand{\eq}[1]{\begin{align} #1 \end{align}}

\begin{document}

\title{Viscosity in the excluded volume hadron gas model}

\author{M.I. Gorenstein}
\affiliation{Bogolyubov Institute for Theoretical Physics, Kiev, Ukraine}
\affiliation{Frankfurt Institute for Advanced Studies, Frankfurt,Germany}

\author{M. Hauer}
\affiliation{Helmholtz Research School, University of Frankfurt, Frankfurt,
  Germany} 

\author{O.N. Moroz}
\affiliation{National Technical University, Kiev, Ukraine}
\affiliation{Bogolyubov Institute for Theoretical Physics, Kiev,
  Ukraine}

\begin{abstract}
The shear viscosity $\eta$ in the van der Waals excluded volume
hadron-resonance gas model is considered. For the shear viscosity the
result of the non-relativistic gas of hard-core particles
is extended to the mixture of particles with different masses, but
equal values of hard-core radius $r$. The relativistic corrections to
hadron average momenta in thermal equilibrium are also taken
into account. The ratio of the viscosity $\eta$ to the entropy
density $s$ is studied. It monotonously decreases along the chemical
freeze-out line in nucleus-nucleus collisions with increasing collision
energy. As a function of hard-core radius $r$, a broad minimum of the ratio
$\eta/s\approx 0.3$ near $r \approx 0.5$~fm is found at high collision
energies. For the charge-neutral system at $T=T_c=180$~MeV, a minimum of the
ratio $\eta/s\cong 0.24$ is reached for $r\cong 0.53$~fm. To justify a
hydrodynamic approach to nucleus-nucleus collisions within the hadron phase
the restriction from below, $r~ \ge ~0.2$~fm, on the hard-core hadron
radius should be fulfilled in the excluded volume hadron-resonance gas. 
\end{abstract}



\maketitle

%-----------------------------------------------------------------------
\section{Introduction}
One of the important discoveries in Au+Au collisions at BNL RHIC
is evidence for strong collective effects in the form of transverse flow
and asymmetric azimuthal (elliptic) flow. To account for this, one needs {\it
  almost perfect fluid} hydrodynamics (i.e. a small ratio, $\eta/s\le 0.2$, of
shear viscosity $\eta$ to entropy density $s$) starting at the very early
stage of nucleus-nucleus (A+A) collision (see, e.g.,
\cite{rhic,rhic1,rhic2,rhic3} and reference therein). Perturbative QCD
calculations \cite{pertur} have led to a much higher value of the viscosity to
entropy density ratio, $\eta/s \ge 1$ (see Ref.~\cite{eta1}). Based on these
results, the `almost perfect fluid' in the deconfined phase is assumed to be 
the strongly interacting quark-gluon plasma (sQGP) \cite{sQGP1}.  In
Ref.~\cite{eta1}, different possibilities for a structure of a minimum 
of the ratio $\eta/s$ at the transition between hadron and quark-gluon phases
are discussed. An absolute minimum of this ratio was identified with the
critical point of QCD matter \cite{eta1,rhic2}. This conclusion has been based
in Ref.~\cite{eta1} on an empirical observation in liquid-gas transitions for
other known substances. In Ref.~\cite{eta2} it was shown that certain field
theories have the ratio $\eta/s=1/4\pi$, and it was conjectured that
$\eta/s\ge 1/4\pi$ for all other substances.  This restriction appears to be 
close to the quantum (due to Heisenberg uncertainty principle)
kinetic lower bound of Ref.~\cite{DG}, $\eta/s \ge 1/15$, 
for the gas of quarks and gluons (see Ref.~\cite{rhic1}).

It is of both practical and  fundamental
importance to know the ratio $\eta/s$ in the hadron phase, in particular,
near the transition to the sQGP.  This will be
a subject of the present paper. We will use the statistical
model with full hadron-resonance spectrum and take into account
the short-range repulsion effects within the hard-core approximation.
Earlier calculations of the ratio $\eta/s$ were done in 
Ref.~\cite{eta4} for the pion gas and pion-kaon mixture (see recent 
pion gas results in Refs.~\cite{etameson1,etapion1,etapion2}). 
There were also the calculations of the hadron system 
viscosity within the microscopic transport models:
UrQMD in Ref.~\cite{transport1} and URASiMA event generator in
Ref.~\cite{transport2}.

During last decades statistical models of hadron-resonance gas
(HG) have served as an important tool to investigate high energy
nuclear collisions. The main object of the study have been 
mean multiplicities of produced hadrons (see, e.g., recent papers
\cite{FOC,FOP,pbm} and references therein). An extension to the 
ideal gas picture based on the van der Waals (VDW) excluded volume procedure
was suggested  in Refs.~\cite{vdw,vdw1,crit} to phenomenologically
take into account repulsive interactions between hadrons. This
leads to the well known VDW suppression of hadron number
densities. However, the particle number ratios are (in Boltzmann
approximation) independent of the proper volume parameter $\upsilon$, if it is
the same for all hadron species. Thus, a rescaling of the total volume $V$
leads to exactly the same hadron yields as those in the ideal hadron gas.
Recently the VDW HG model has been used in Ref.~\cite{vdw-fluc} to
calculate particle number fluctuations. It was demonstrated
\cite{vdw-fluc} that multiplicity fluctuations are suppressed
in the VDW HG gas. This suppression is qualitatively different
from that of particle yields. In contrast to average
multiplicities, the suppression of multiplicity fluctuations cannot be removed
by rescaling of the total volume of the system.  Thus, the hard-core radius of
the hadrons can be, in principle, straightforwardly connected with  data on
multiplicity fluctuations. 

The aim of the present paper is to make  calculations of the
ratio $\eta/s$ in the VDW HG model. We present our results for different
values of temperature $T$ and baryon chemical potential
$\mu_B$ along the chemical freeze-out line. In this $T-\mu_B$ region
 hydrodynamic expansion within the hadron phase is expected.  The paper is
organized as follows. In Section II we present the results for the
thermodynamic functions in the multi-component VDW gas. In Section III the
expression for the viscosity in the VDW HG is obtained. In Section IV the
viscosity to entropy density ratio is calculated along the chemical freeze-out
line. This gives the VDW HG predictions for this ratio in central
nucleus-nucleus collisions  in the hadron phase at different collision
energies. A summary, presented in Section V, closes the paper. 

\section{VDW Hadron Gas }

The VDW excluded volume procedure gives the following
transcendental equation for the system pressure $p$  \cite{vdw}:
\eq{ \label{vdwp}
  p~=~\exp\left(-~\frac{\upsilon~ p}{T}\right)~ T~\phi ~,
}
where $\upsilon$ is the hadron excluded volume parameter,
\eq{\label{v}
  \upsilon~=~4\cdot \frac{4}{3}\pi~ r^3~,
}
with $r$ being the particle hard-core radius. In what follows it 
is assumed to be the same for all hadron species. In Eq.~(\ref{vdwp}), $\phi$
is the total particle number density in the ideal HG:
\eq{\label{phi}
  \phi(T,\mu_B)~=~\sum_i \phi_i~=~\sum_i \frac{g_i}{2\pi^2}~ m_i^2T
  K_2\left(\frac{m_i}{T}\right) \exp\left(\frac{\mu_i}{T}\right)~,
}
with $\phi_i$ in Eq.~(\ref{phi}) being the ideal gas particle
number density of $i$th particle species.  As in Ref.~\cite{vdw-fluc}, we use
the Boltzmann approximation, neglecting small effects of Bose and/or Fermi
statistics. In Eqs.~(\ref{vdwp},\ref{phi}), $T$ is the system temperature,
$g_i$ the degeneracy factor of $i$th particle species, $m_i$ the particle
mass, and $\mu_i=b_i\mu_B+s_i\mu_S+q_i\mu_Q$ the chemical potential due to the
$i$th particle charges $(b_i,s_i,q_i)~ -$ baryon number, strangeness, electric
charge. The summation in Eq.~(\ref{phi}) is taken over all hadron
species. Finally, $K_2$ in Eq.~(\ref{phi}) is the modified Hankel
function. The $i$th particle number density in the VDW HG equals to:
\eq{\label{ni}
  n_i~=~\frac{ \exp(-~\upsilon p/T)~\phi_i}{ 1~+~\upsilon ~\exp(-~\upsilon
    p/T)~\phi}~\equiv~ \frac{x_i}{1~+~\upsilon~x}~.
}
In Eq.~(\ref{ni}) the notations, $x_i\equiv \exp(-\upsilon p/T)\phi_i$ and
$x\equiv \sum_ix_i$, have been introduced. The factor $R=\exp(-\upsilon
p/T)~(1+\upsilon~x)^{-1}$ presents the VDW suppression factor.  Since the same
excluded volume parameter for all particle species is assumes, it is the same
for all particle densities, and the VDW HG energy density.  

The VDW HG entropy density $s$ is equal to:
\eq{
  \label{s}
  s~=~\frac{\partial p}{\partial T}~=~ \frac{x}{1~+~\upsilon x}~\left(
    1~+~\upsilon x~+~\frac{T}{\phi}~\frac{\partial \phi}{\partial T}\right).
}

\section{shear viscosity}
The shear viscosity in a non-relativistic gas of hard-core
spheres can be calculated analytically (see e.g.,
Ref.~\cite{visc}):
 \eq{ \label{eta1}
 \eta~=~\frac{5}{64~\sqrt{\pi}}~\frac{\sqrt{m~T}}{r^2}~.
 }
The Eq.~(\ref{eta1}) demonstrates a simple dependence of the
viscosity on the system temperature $T$, particle mass $m$, and
hard-core particle radius $r$. The derivation of Eq.~(\ref{eta1})
from the molecular-kinetic theory demonstrates the dependence of the
viscosity, 
\eq{\label{eta0}
  \eta~\propto~n~l~\langle |{\bf p}| \rangle~,
}
on the particle density  $n$, the mean free path $l\propto 1/(nr^2)$, and the
particle average thermal momentum:
\eq{\label{p-nonrel}
  \langle | {\bf p} | \rangle~=  ~\frac{\int_0^{\infty}p^2dp~p~
    \exp[-~p^2/(2mT)]}{\int_0^{\infty}p^2dp~\exp[-~p^2/(2mT)]}~=~  
  \sqrt{\frac{8mT}{\pi}}~.
}

The Eq.~(\ref{p-nonrel}) can be easily extended  to the relativistic
thermal motion:
\eq{\label{p-rel}
 \langle | {\bf p} |
 \rangle~=~\frac{\int_0^{\infty}p^2dp~p~\exp[-~\sqrt{p^2+m^2}/T]} 
 {\int_0^{\infty}p^2dp~\exp[-~\sqrt{p^2+m^2}/T]}~=~ 
 \sqrt{\frac{8mT}{\pi}}~\frac{K_{5/2}(m/T)}{K_2(m/T)}~.
}
At $m/T \gg 1$, the factor $K_{5/2}(m/T)/K_2(m/T)$ goes to 1, and
Eq.~(\ref{p-rel}) is reduced to Eq.~(\ref{p-nonrel}). In the opposite limit,
$m/T \ll 1$, Eq.~(\ref{p-rel}) goes to the well  known result for massless
particles, $\langle | {\bf p} | \rangle = 3T$. 

For the the mixture of particle species with different masses, but
with the same hard-core radius $r$, Eq.~(\ref{eta1}) is transformed into:
\eq{
  \label{eta2} \eta=\frac{5}{64~\sqrt{\pi}}~\frac{\sqrt{T}}{r^2}~
  \sum_{i=1}\sqrt{m_i}~\frac{K_{5/2}(m_i/T)}{K_2(m_i/T)}~\frac{n_i}{n}~,
}
where $n_i$ is given by Eq.~(\ref{ni}), and $n=\sum_i n_i$ is the
total VDW gas particle number density. In Eq.~(\ref{eta2}), the
relativistic correction factors, $K_i\equiv K_{5/2}(m_i)/T)/K_2(m_i/T)$ from
Eq.~(\ref{p-rel})), are taken into account. To obtain Eq.~(\ref{eta2}) the
same mean free path has been assumed for different $i$th hadron species.
This is approximately valid because of the same hard-core radius $r$.   
The corrections because of different particle masses $m_i$ are neglected. 
Eq.~(\ref{eta2}), with the summation over all particles  and resonance species
listed in THERMUS \cite{Thermus}, will be used to calculate the hadron
viscosity. The ratios $n_i/n$ in Eq.~(\ref{eta2}) can be substituted by the
ideal gas values, $n_i/n=\phi_i/\phi$, as we have assumed the same hard-core
radius $r$ for all hadrons. The VDW suppression factor $R$ is then the same
for all hadron species, and it is canceled out in the ratios of particle
densities.  

Note that the viscosity $\eta$ is not proportional to the number of hadron
species. If one assumes the same hadron masses $m_i=m$ and $m\gg T$,  
Eq.~(\ref{eta2}) is reduced back to the case of one particle species
(\ref{eta1}). On the other hand, the entropy density $s$ increases with the
number of species. This fact has been used in recent paper \cite{example} 
to construct a counterexample to the lower bound of $\eta/s=1/4\pi$ assumed in
Ref.~\cite{eta2}. 

In the limit $r\rightarrow 0$ the excluded volume parameter
$\upsilon$ (\ref{v}) goes to zero. In this limit, the
thermodynamical functions of the VDW HG, $p$ (\ref{vdwp}),
$n_i$ (\ref{ni}), and $s$ (\ref{s}) converge to the corresponding
ideal gas expressions, $p^{id}=T\phi$, $n_i^{id}=\phi_i$, 
and $s^{id}= \phi+ T\partial \phi/\partial T$.
On the other hand, according to
Eq.~(\ref{eta2}) the viscosity $\eta$  goes to infinity as $\eta\propto 1/r^2$.
Thus, the ratio $\eta/s$ diverges as $r^{-2}\rightarrow\infty$  in the
ideal gas limit of $r\rightarrow 0$. 

\section{VDW HG Model Results}
In this section we present the results of the VDW HG for the ratio $\eta/s$
along the chemical freeze-out line in central  A+A  collisions for the whole
energy range from SIS to LHC. The values of $T$, $\mu_B$, and $\gamma_s$ at
the chemical freeze-out at different collision energies are presented in Table
I. They are identical to those values in Table~I of Ref.~\cite{vdw-fluc}.  
\begin{table}[h!]
\begin{center}
\begin{tabular}{||c||c|c|c||c||c|c||c|c||c|c||c||}\hline
      & &  & & &
   \multicolumn{2}{c||}{VDW HG} & 
    \multicolumn{2}{c||}{VDW HG} & 
   \multicolumn{2}{c||}{VDW HG} \\
 $\sqrt{s_{NN}}$ & $T$ & $\mu_B$ & $\gamma_S$ & $K_{\pi}$ & 
   \multicolumn{2}{c||}{ ${\bf r}$ {\bf =~0.1~fm}} & 
   \multicolumn{2}{c||}{ ${\bf r}$ {\bf =~0.3~fm}} & 
   \multicolumn{2}{c||}{ ${\bf r}$ {\bf =~0.5~fm}} 
   \\ [0.5ex]
\hline [ GeV ] &[ MeV ] &[ MeV ] & & & $R$ & $\eta/s$ & $R$ & $\eta/s$ & $R$ & $\eta/s$ 
 \\
[0.5ex] \hline\hline
$ 2.32 $ & 64.3  & 800.8 & 0.64 & 1.50 & 0.998 & 16.6 & 0.944 & 1.94 & 0.788 &0.82 \\
$ 4.86 $ & 116.5 & 562.2 & 0.69 & 1.85 & 0.994 & 6.92 & 0.870 & 0.87 & 0.603 &0.44 \\
$ 6.27 $ & 128.5 & 482.4 & 0.72 & 1.92 & 0.993 & 5.68 & 0.844 & 0.73 & 0.552 &0.39 \\
$ 7.62 $ & 136.1 & 424.6 & 0.73 & 1.97 & 0.992 & 5.01 & 0.825 & 0.66 & 0.519 &0.37 \\
$ 8.77 $ & 140.6 & 385.4 & 0.75 & 1.99 & 0.991 & 4.64 & 0.812 & 0.62 & 0.498 &0.35 \\
$ 12.3 $ & 149.0 & 300.1 & 0.79 & 2.04 & 0.990 & 4.03 & 0.786 & 0.55 & 0.459 &0.33 \\
$ 17.3 $ & 154.4 & 228.6 & 0.83 & 2.07 & 0.989 & 3.66 & 0.766 & 0.51 & 0.432 &0.31 \\
$ 62.4 $ & 160.6 & 72.5  & 0.98 & 2.11 & 0.987 & 3.18 & 0.738 & 0.46 & 0.397 &0.30 \\
$ 130  $ & 161.0 & 35.8  & 1.0  & 2.11 & 0.986 & 3.14 & 0.735 & 0.46 & 0.393 &0.29 \\
$ 200  $ & 161.1 & 23.5  & 1.0  & 2.11 & 0.986 & 3.13 & 0.735 & 0.46 & 0.393 &0.29 \\
$ 5500 $ & 161.2 & 0.9   & 1.0  & 2.11 & 0.986 & 3.13 & 0.735 & 0.46 & 0.393 &0.29 \\
\hline
\end{tabular}
\label{Table1}
\caption{The chemical freeze-out parameters $T$, $\mu_B$, and
$\gamma_S$  in central heavy-ion collisions are presented at different
c.m. energies $\sqrt{s_{NN}}$. The  ratio $\eta/s$ is presented in the VDW HG
for three values of the hard-core radius, $r=0.1$~fm,~0.3~fm, and 0.5~fm. 
The VDW suppression factor, $R=\exp(-\upsilon p/T)~(1+\upsilon x)^{-1}$,  for
these hard-core radiuses  is also presented at different freeze-out points.
The factor $K_{\pi}=K_{5/2}(m_{\pi}/T)/K_2(m_{\pi}/T)$ shows the relativistic
enhancement for the pion thermal momentum at temperature $T$ according to
Eq.~(\ref{p-rel}). 
}
\end{center}
\end{table}

Note that the conditions for average energy per particle, $\langle
E\rangle/\langle N\rangle =1$~GeV \cite{Cl-Red},  zero value of the net total 
strangeness, $S=0$ (this defines $\mu_S=\mu_S(T,\mu_B)$), and the charge to
baryon ratio, $Q/B = 0.4$ (this defines $\mu_Q=\mu_Q(T,\mu_B)$), remain the
same as in Refs.~\cite{res_CE,res_MCE}. The only tiny difference of $T$ and
$\mu_B$ values from those in Refs.~\cite{res_CE,res_MCE} comes because of the
Boltzmann statistics approximation used in the present paper and in
Ref.~\cite{vdw-fluc}. The dependence of $\mu_B$ on the collision energy is
parameterized as \cite{FOC}, $\mu_B \left( \sqrt{s_{NN}} \right)
=1.308~\mbox{GeV}~(1+ 0.273~ \sqrt{s_{NN}})^{-1}~$, where the
c.m. nucleon-nucleon collision energy, $\sqrt{s_{NN}}$, is taken in GeV
units. The strangeness saturation factor, $\gamma_S$, is  parameterized as
\cite{FOP}, $ \gamma_S~ =~ 1 - 0.396~ \exp \left( - ~1.23~ T/\mu_B \right)$.
Both these relations are the same as in Refs.~\cite{vdw-fluc,res_CE,res_MCE}.
The resulting ratio $\eta/s$ in the VDW HG is presented in Fig.~1 as a
function of hard-core radius $r$ for several chemical freeze-out points from
Table I. 

\begin{figure}[h!]
 \begin{center}
 \epsfig{file=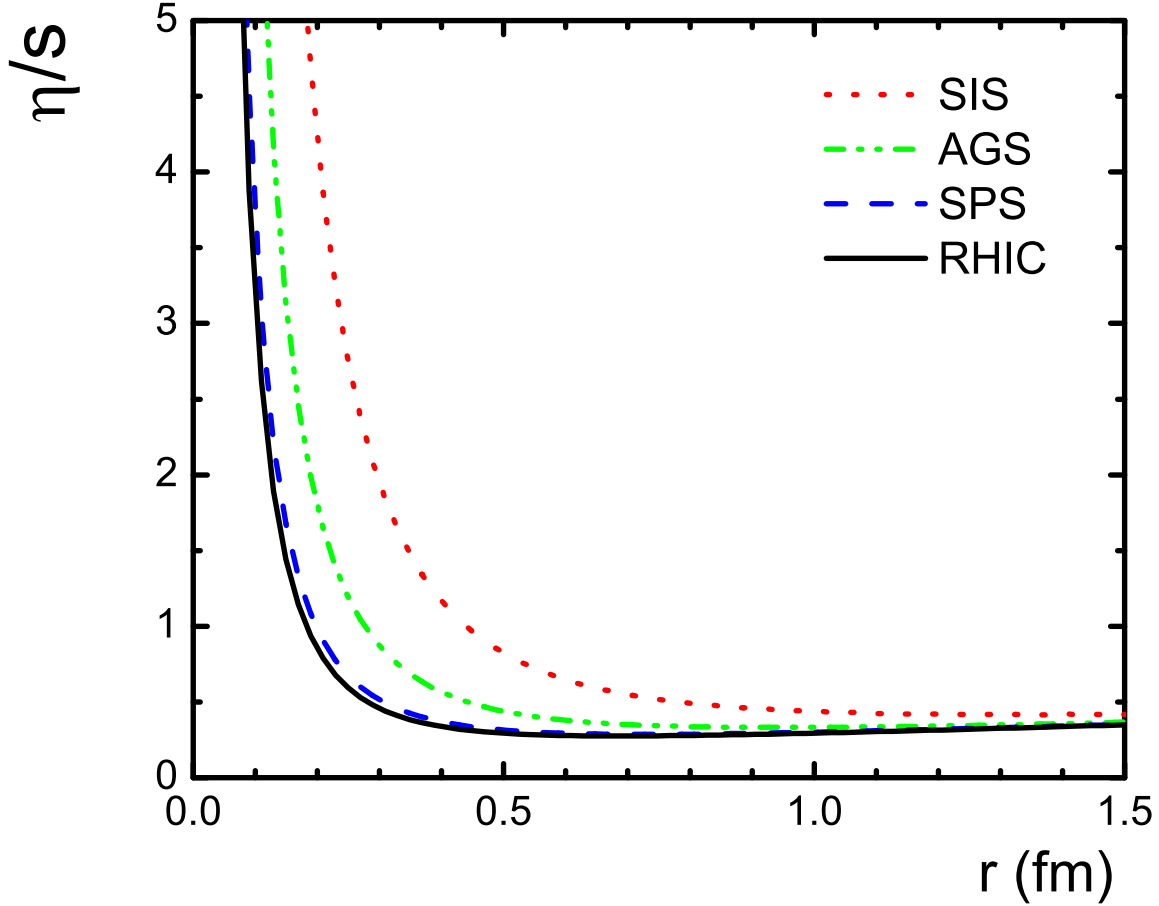,width=16cm}
 \caption{The ratio $\eta/s$ in the VDW HG
 as the function of hard-core radius $r$.
 The lines correspond to different $T-\mu_B$ freeze-out points from Table I:
 SIS ($\sqrt{s_{NN}}=2.32$~GeV), AGS ($\sqrt{s_{NN}}=4.86$~GeV), SPS
 ($\sqrt{s_{NN}}=17.3$~GeV), and RHIC ($\sqrt{s_{NN}}=200$~GeV). \label{rdep}
} 
\end{center}
\end{figure}

\begin{figure}[h!]
 \begin{center}
 \epsfig{file=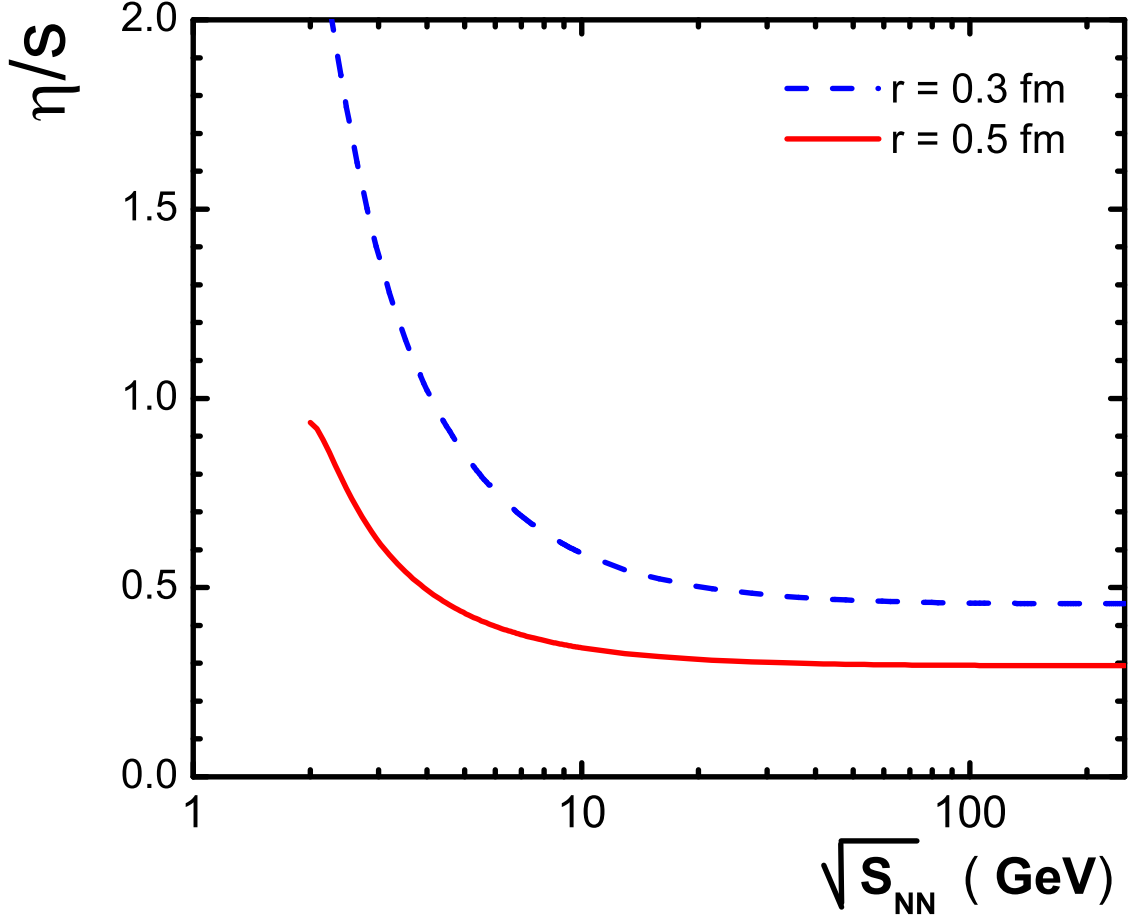,width=16cm}
 \caption{The ratio $\eta/s$ in the VDW HG along the chemical freeze-out line 
as the function of $\sqrt{s_{NN}}$ at different hard-core radiuses. The solid
line presents the results for $r$=0.5~fm and dashed line for $r$=0.3~fm.
\label{sdep} }
\end{center}
\end{figure}

The values of $\sqrt{s_{NN}}$ quoted in Table~I correspond to the beam
energies at SIS (2$A$~GeV), AGS (11.6$A$~GeV), SPS ($20A$, $30A$, $40A$,
$80A$, and $158A$~GeV), colliding energies at RHIC ($\sqrt{s_{NN}}=62.4$~GeV,
$130$~GeV, and $200$~GeV), and LHC ($\sqrt{s_{NN}}=5500$~GeV). The excluded
volume correction factor, $R= \exp \left( - \upsilon p/T \right)~\left( 1+
  \upsilon x \right)^{-1}$, and the entropy density, Eq.~(\ref{s}), are
calculated using the THERMUS package \cite{Thermus}. The relativistic
enhancement factor, $K_{\pi}=K_{5/2}(m_{\pi}/T)/K_2(m_{\pi}/T)$, for 
average thermal momentum is also presented in Table I at different freeze-out
points for pions, which are the lightest hadrons. The viscosity is calculated
according to Eq.~(\ref{eta2}). The Fig.~\ref{rdep} demonstrates a presence of
a broad minimum of the ratio $\eta/s$ near $r \approx 0.5$~fm at high
collision energies. For fixed hard-core radius $r$, the ratio $\eta/s$
monotonously decreases along the freeze-out line with increasing of collision
energy. This is shown in Fig.~\ref{sdep}.   

The presence of strong collective flow in A+A collisions  indicates a rather
small viscosity. According to Ref.~\cite{visc-disc} the following inequality,
$\eta/s~\le~ 5$~, should be satisfied. As seen from Fig.~\ref{rdep}, this
leads to the restriction, $r~\ge~0.2~\mbox{fm}$~, on the hard-core radius from
below in the VDW HG. 

\begin{figure}[h!]
 \begin{center}
   \epsfig{file=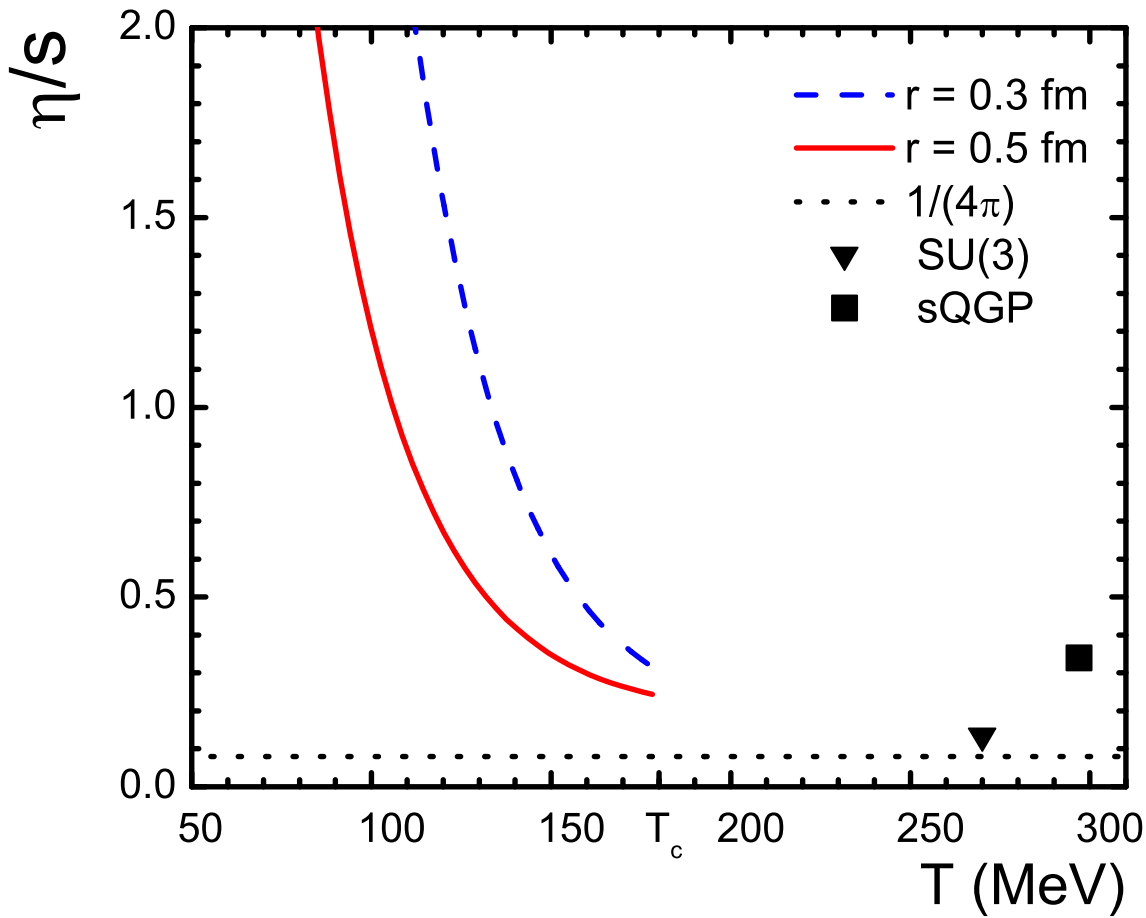,width=16cm}
   \caption{The ratio $\eta/s$ for charge-neutral system ($Q=B=S=0$ and
     $\gamma_S=1$) as a function of $T$. We fix the crossover temperature
     between HG and sQGP as $T_c=180$~MeV. At $T\le T_c$ the VDW HD results
     are shown at different hard-core radiuses, the solid line corresponds to 
     $r$=0.5~fm  and dashed line to $r$=0.3~fm. The dotted line corresponds to
     the lower bound, $\eta/s =1/4\pi$ \cite{eta2}. At $T> T_c$, the triangle
     symbol corresponds to Ref.~\cite{visc-lQCD2} of the pure gluon plasma in
     the lattice QCD, and the square is the estimate of  Ref.~\cite{visc-sQGP}
     in the quasiparticle model of the sQGP.
     \label{Tdep} }
 \end{center}
\end{figure}

The Fig.~\ref{Tdep} shows the dependence of the ratio $\eta/s$ on the
temperature $T$ in the VDW HG for charge-neutral system  ($Q=B=S=0$ and
$\gamma_S=1$).  

The VDW HG for $r=$0.3~fm and $r$=0.5~fm  are presented at $T<T_c$, with
$T_c=180$~MeV assumed as the temperature of a crossover  between the confined
and deconfined phase. At $T=T_c=180$~MeV a minimum of the ratio $\eta/s\cong
0.24$ is reached for $r\cong 0.53$~fm. Note that at high collision energies
the baryon chemical potential is small, and a hydrodynamic expansion within
the hadron phase at $T<T_c$ is expected in A+A collisions.

In the deconfined phase there were several estimates of the ratio
$\eta/s$.  The estimates of Ref.~\cite{eta1} based on the perturbative
calculations \cite{pertur}  give the large value, $\eta/s>1$. 
The non-perturbative results correspond to smaller values
of the ratio $\eta/s$. In the pure SU(3) gauge model on the lattice a shear
viscosity to entropy density ratio of $\eta/s <1 $ in the temperature region
$1.4\le T/T_c \le 1.8 $ was found in Ref.~\cite{visc-lQCD1} . The same upper
bound, $\eta/s <1 $,  and the estimate $\eta/s=0.134\pm 0.033$ at $T=1.65~T_c$
were obtained in Ref.~\cite{visc-lQCD2}. In the quasiparticle description of
the sQGP it was found \cite{visc-sQGP} a shear viscosity to entropy density
ratio $\eta/s \approx 0.34$ at $T\approx 1.5T_c$. 

\section{summary}
In summary, the shear viscosity $\eta$ in the van der Waals excluded volume
hadron-resonance gas model is calculated. We use the same hard-core radius $r$
for all hadron species. The ideal gas limit appears to be a singular one: the
ratio of the viscosity $\eta$ to the entropy density $s$ diverges as $r^{-2}$
at $r\rightarrow 0$. This makes an ideal gas picture inappropriate 
for any kinetic or hydrodynamic calculations. Moreover, the ratio $\eta/s$
should be small enough to justify the hydrodynamic approach to A+A
collisions. Near the chemical freeze-out line this leads to the restriction
from below, $r~ \ge ~0.2$~fm, on the hard-core hadron radius. When the ratio
$\eta/s$ is considered as a function of $r$, a broad minimum, $\eta/s \approx
0.3$~fm for $r \approx 0.5$~fm is found for the chemical freeze-out at high
collision energies. For fixed hard-core radius $r$, the ratio $\eta/s$
monotonously decreases along the chemical freeze-out line with increasing 
collision energy. A similar behavior is found for the ratio $\eta/s$ as a
function of $T$ for the charge-neutral system. The ratio $\eta/s$ decreases
with $T$, and its minimal value is $\eta/s\cong  0.24$ at $T=T_c=180$~MeV for
$r\cong 0.53$~fm. Theoretical estimates of the ratio $\eta/s$ in the
deconfined phase require the non-perturbative calculations, and they are far
from being complete. More efforts are needed to clarify the detailed behavior
of the ratio $\eta/s$ at the transition (crossover) between hadron and
quark-gluon phases.   

\begin{acknowledgments}
We would like to thank  V.V. Begun, W.~Greiner, O.A. Mogilevsky,
S.~Mrowczynski, G.~Torrieri and O.S.~Zozulya for discussions. One of the
author (M.I.G.) is thankful to the FIAS and ITP for the warm hospitality in
Frankfurt. 
\end{acknowledgments}


\begin{thebibliography}{100}

\bibitem{rhic}
D. Teaney, Phys. Rev. C {\bf 68}, 034913 (2003).

\bibitem{rhic1}
T. Hirano and M. Gyulassy, Nucl. Phys. A {\bf 769}, 71 (2006).

\bibitem{rhic2}
R.A. Lacey, et al. Phys. Rev. Lett. {\bf 98}, 092301 (2007).

\bibitem{rhic3}
H. Drescher, A. Dumitru, C. Gombeaud, and J. Ollitrault,
arXiv:nucl-th/0704.3553. 

\bibitem{pertur}
P. Arnold, G.D. Moore, and L.G. Yaffe, J. High Energy Physics, 05
(2003) 051.

\bibitem{eta1}
L.P. Csernai, J.I. Kapusta, and L.D. McLerran, Phys. Rev. Lett.
{\bf 97} 152303 (2006).

\bibitem{sQGP1}
T.D. Lee, Nucl. Phys. A {\bf 750}, 1 (2005); M. Gyulassy and L.
McLerran,  Nucl. Phys. A {\bf 750}, 30 (2005); E.V. Shuryak, Nucl.
Phys. A {\bf 750}, 64 (2005).

\bibitem{eta2} P.K. Kovtun, D.T. Son, and A.O. Starinets, Phys. Rev. Lett.
{\bf 94} 111601 (2005).

\bibitem{DG}
P. Danielewicz and M. Gyulassy, Phys. Rev. D {\bf 31}, 53 (1985).

\bibitem{eta4}
M. Prakash, M. Prakash, R. Venugopalan, and G. Welke, Phys. Rep.
{\bf 227}, 321 (1993).

\bibitem{etameson1}
A. Dobado and F.J. Llanes-Estrada, Phys. Rev. D {\bf 69}, 116004
(2004).

\bibitem{etapion1}
J. Chen and E. Nakano, Phys. Lett. B {\bf 647}, 371 (2007).

\bibitem{etapion2}
E. Nakano, arXiv:hep-ph/0612255.

\bibitem{transport1}
A. Muronga, Phys. Rev. C {\bf 69}, 044901 (2004).

\bibitem{transport2}
S. Muroya and N. Sasaki, Prog. Theor. Phys. {\bf 113}, 457 (2005).

\bibitem{FOC} J. Cleymans, H. Oeschler, K. Redlich,
  and S. Wheaton, Phys. Rev. C {\bf 73}, 034905 (2006).

\bibitem{FOP} F. Becattini, J. Manninen, and M. Ga\'zdzicki,
  Phys. Rev. C {\bf 73}, 044905 (2006).

\bibitem{pbm}
A. Andronic, P. Braun-Munzinger, J. Stachel, Nucl. Phys. A {\bf
772}, 167 (2006).

 \bibitem{vdw}
D.H. Rischke, M.I. Gorenstein, H. St\"ocker, and W. Greiner, Z.
Phys. C {\bf 51}, 485 (1991);\\
J. Cleymans, M.I. Gorenstein, J. Stalnacke, and E.~Suhonen, Z.
Phys. C {\bf 8}, 347 (1993).

\bibitem{vdw1}
Granddon D. Yen, M.I. Gorenstein, W. Greiner, and Shin Nan Yang,
Phys. Rev. C {\bf 56}, 2210 (1997);
M.I. Gorenstein, H. St\"ocker, Granddon D. Yen, Shin Nan Yang, and
W. Greiner, J. Phys. G {\bf 24}, 1777 (1998);
Granddon D. Yen and M.I. Gorenstein, Phys. Rev. C {\bf 59}, 2788
(1999).

\bibitem{crit}
M.I. Gorenstein, V.K. Petrov, and G.M. Zinovjev, Phys. Lett. B
{\bf 106}, 327 (1981);
M.I. Gorenstein, W. Greiner, and Shin Nan Yang, J.Phys. G {\bf
24}, 725 (1998); M.I. Gorenstein, M. Ga\'zdzicki, and W. Greiner,
Phys. Rev. C {\bf 72}, 024909 (2005).

\bibitem{vdw-fluc}
M.I. Gorenstein, M. Hauer, and D.O. Nikolajenko, arXiv:nucl-th/0702081,
Phys. Rev. C, in print.

\bibitem{visc}
E. M. Lifschitz and L. P. Pitaevski, Physical kinetics, 2.
ed. Pergamon Press, Oxford, 1981 (Landau-Lifschitz. Course of theoretical
physics. V.10), Chap. 1, Par. 10.

\bibitem{Thermus} 
  S.~Wheaton and J.~Cleymans, arXiv:hep-ph/0407174.
 
\bibitem{example}
T.C. Cohen, Phys. Rev. Lett. {\bf 99}, 021602 (2007).

\bibitem{Cl-Red} J. Cleymans and K. Redlich, Phys. Rev. Lett. {\bf
81}, 5284 (1998).

\bibitem{res_CE}
V.V. Begun, M.I. Gorenstein, M. Hauer, V.P. Konchakovski, and O.S.
Zozulya, Phys.Rev. C {\bf 74} 044903 (2006).

\bibitem{res_MCE}
V.V. Begun, M. Ga\'zdzicki, M.I. Gorenstein, M. Hauer, V.P.
Konchakovski, and B. Lungwitz arXiv:nucl-th/0611075, Phys. Rev. C, in print.

\bibitem{visc-disc} L.P. Csernai, J.I. Kapusta, and L.D. McLerran,
J. Phys. G {\bf 32} S115 (2006).

\bibitem{visc-lQCD1}
A. Nakamura and S. Sakai, Phys. Rev. Lett. {\bf 94}, 072305
(2005).

\bibitem{visc-lQCD2}
H.B. Meyer, arXiv:hep-lat/0704.1801.

\bibitem{visc-sQGP}
B.A. Gelman, E.V. Shuryak, and I. Zahed, Phys. Rev. C {\bf 74},
044908 (2006).

\end{thebibliography}
\end{document}